\documentclass[aps,prb,twocolumn,amsmath,amssymb,footinbib,showpacs]{revtex4}

\usepackage{graphicx}
\usepackage{amsmath}
\usepackage{enumitem}
\usepackage{natbib}
\usepackage{float}
\usepackage{verbatim}
\usepackage{bm}
\usepackage{color}

\begin{document}

\title{Quasiclassical theory of disordered Rashba superconductors}

\author{Manuel Houzet}
\author{Julia S. Meyer}
\affiliation{Univ.~Grenoble Alpes, INAC-SPSMS, F-38000 Grenoble, France}
\affiliation{CEA, INAC-SPSMS, F-38000 Grenoble, France}

\begin{abstract}
We derive the quasiclassical equations that describe two-dimensional superconductors with a large Rashba spin-orbit coupling and in the presence of impurities. These equations account for the helical phase induced by an in-plane magnetic field, with a superconducting order parameter that is spatially modulated along a direction perpendicular to the field. We also derive the generalized Ginzburg-Landau functional, which includes a linear-in-gradient term corresponding to the helical phase. This theory paves the way for studies of the proximity effect in two-dimensional electron gases with large spin-orbit coupling.
\end{abstract}

\pacs{74.78.-w, 75.70.Tj}


\maketitle

\section{Introduction}
Breaking time-reversal symmetry in superconductors that lack inversion symmetry can result in new and interesting phenomena, such as topological phases~\cite{BookTopoIns} or magneto-electric couplings~\cite{BookNCS}. For instance, the application of an in-plane magnetic field to a non-centrosymmetric superconducting layer allows for the formation of a ``helical'' phase, which is characterized by a spontaneously modulated order parameter in the direction transverse to the field~\cite{Edelstein1989}. On the phenomenological level, this helical phase results from the fact that the Ginzburg-Landau functional may admit a linear-in-gradient term~\cite{Mineev1994}. This term, also known as a Lifshitz invariant, describes the coupling between the magnetic field and the supercurrent in the presence of spin-orbit coupling. Thus, the helical state is conceptually different from the nonuniform Fulde-Ferrell-Larkin-Ovchinnikov (FFLO) state~\cite{Fulde,Larkin}, which arises from a change of sign of the quadratic-in-gradient term~\cite{Buzdin1997} that fixes the amplitude of the modulation, but not its direction.

The amplitude of the Lifshitz invariant was calculated for the microscopic model of a two-dimensional superconductor with Rashba coupling in the clean limit~\cite{Edelstein1996}. This study showed that the modulation wavevector in the helical phase is proportional to the spin-orbit-induced difference in the densities of state of electrons with opposite helicities. The parallel upper critical field in the helical phase was  computed both in the clean~{ \cite{Barzykin2002,Loder2013}} and diffusive~\cite{Dimitrova2003} regimes. The relevance of this effect for superconductivity at oxide interfaces~\cite{Reyren1997,Michaeli2012} and in single-atomic layers of Pb~\cite{Sekihara2013} was also discussed. Furthermore, the possibility of a helical phase in three-dimensional superconductors was debated~\cite{Kaur2005,Samokhin2008,MineevSigrist}.

Here we provide a microscopic theory of the helical phase in the disordered case.
{ While homogeneous superconductivity is basically unaffected by disorder as long as the localization length remains much larger than the coherence length (Anderson theorem \cite{AG58,Anderson59}), the FFLO state is rapidly destroyed by even weak disorder~\cite{Aslamazov68}. By contrast, the helical phase is modified by the disorder, but it survives up to fairly large disorder strengths.}
The quasiclassical theory offers a convenient framework to discuss the equilibrium and out-of-equilibrium properties of superconductors~\cite{BookKopnin}. The difficulty to elaborate such a theory in the case of non-centrosymmetric superconductors is related with the large spin-orbit induced splitting, which is responsible for the helical phase. Namely, the normal-state dispersion of the material consists of two so-called helical bands with different densities of state at the Fermi level.
In the clean limit, a large spin-orbit coupling results in an effective two-band superconductivity with Cooper pairs that form separately in each of the helical bands, while sharing the same gap function~\cite{MineevSamokhin}. The paramagnetic effect induced by the magnetic field results in a modulation of the order parameter similar to the FFLO phase~\cite{Fulde,Larkin}. 
{ The preferred modulation wavevector is opposite in the two bands. Thus, the modulation results from a competition between the two bands and their difference in densities of states is crucial in order to obtain a non-vanishing result.} Quasiclassical Eilenberger equations have been derived to describe this situation \cite{Agterberg2007}. Their solution permitted to consider the crossover between FFLO and helical phases as well as to predict the density of states in the helical phase. However, impurity scattering mixes the bands. { Thus, inter-band superconducting correlations may develop.} So far quasiclassical equations for the diffusive regime were derived in the spin basis but neglecting the spin-orbit splitting~\cite{Malshukov2008,Bergeret2013}. Thus, they did not account for the helical phase. More recent works considered the case of very weak spin-orbit coupling \cite{Konschelle2014,Bergeret2014}, where only a higher order effect is expected to survive.

In our work, we derive quasiclassical equations in the helical basis which account for the spin-orbit splitting and therefore describe the helical phase. To set the stage, we first derive the quasiclassical equations in the helical basis when the spin-orbit coupling is small and the helical modulation may be neglected. Within that approximation, we demonstrate that for moderate disorder, despite inter-band impurity scattering, Cooper pairing predominantly occurs in separate bands. Thus, we may derive a set of quasiclassical equations for each helical band separately. This allows us to take into account  the spin-orbit splitting and obtain the helical phase. { The equations we derive describe the entire superconducting phase at arbitrary temperatures and magnetic fields.} We also derive the dependence on disorder of the Lifshitz invariant in the generalized Ginzburg-Landau functional to further characterize the transition to the helical phase.

\section{Model}

A two-dimensional electron gas with Rashba spin-orbit coupling is described by the Hamiltonian
\begin{equation}
\label{eq:HK}
H_0=\sum_{\bf k}a_{\bf k}^\dagger\left[\xi_{\bf k}+\alpha({\bf k}\times\hat{ \bf z})\cdot{\bm \sigma}\right]a_{\bf k}.
\end{equation}
Here, $a_{\bf k}=(a_{{\bf k}\uparrow},a_{{\bf k}\downarrow})^T$ are annihilation operators for an electron with spin $\sigma=\uparrow, \downarrow$ and momentum $\bf k$, $\xi_{\bf k}={\bf k}^2/(2m)-\mu$ is
the kinetic energy associated with the effective mass $m$ and chemical potential $\mu$, $\alpha=\Delta_{\rm so}/\sqrt{2m\mu}$ characterizes the spin-orbit coupling strength, and $\hat{ \bf z}$ is a unit vector normal to the electron gas. Furthermore, Pauli matrices $\sigma_i$ act in spin space. 

The Hamiltonian \eqref{eq:HK} can be diagonalized in the so-called helical basis,
\begin{equation}
\label{eq:HK-helical}
H_0=\sum_{\bf k}c_{\bf k}^\dagger\left(\xi_{\bf k}+\alpha k\sigma_z\right)c_{\bf k},
\end{equation}
where $c_{\bf k}=U_{\bf k}^{-1}a_{\bf k}$, with 
$
U_{\bf k}=
\frac 1{\sqrt{2}}\left(\begin{array}{cc}
1&1\\
-ie^{i{\varphi_{\bf k}}}&ie^{i{\varphi_{\bf k}}}
\end{array}\right).
$
Here, $\varphi_{\bf k}$ is the angle of ${\bf k}$ in the $(xy)$-plane.
According to Eq.~\eqref{eq:HK-helical}, the helical bands are characterized by the same Fermi velocity, $v=\sqrt{2\mu/m+\alpha^2}$, and different densities of states, $\nu_\lambda=\nu(1-\lambda \alpha/v)$, where $\nu=m/(2\pi)$ and $\lambda=\pm$ labels the helical bands.  

A magnetic field $\bf B$ acting on the electron spin induces a Zeeman term,
\begin{equation}
H_Z=\sum_{\bf k}a_{\bf k}^\dagger{\bf h}\cdot{\bm \sigma} a_{\bf k}=\sum_{\bf k}c_{\bf k}^\dagger {\bf h}_{\bf k}\cdot{\bm \sigma}
c_{\bf k},
\end{equation}
in the spin and helical basis, respectively. Here, ${\bf h}=\frac 1 2 g\mu_B{\bf B}$ and ${\bf h}_{\bf k}\equiv(
h_z, -{\bf h}_\parallel\cdot\hat{\bf k},-({\bf h}_\parallel\times\hat{\bf z})\cdot\hat{\bf k})^T$,
where ${\bf h}_\parallel$ and $h_z$ are the in-plane and out-of-plane components of ${\bf h}$, respectively, while $\hat{\bf k}={\bf k}/k$.

Superconductivity is described by a BCS Hamiltonian with s-wave pairing{~\cite{footnote-swave}}, 
\begin{equation}
\label{eq:BCS}
H_S=- V_S\sum_{\bf k,k',q}a_{\bf k\uparrow}^\dagger a_{\bf -k+q\downarrow}^\dagger a_{\bf -k'+q\downarrow} a_{\bf k'\uparrow},
\end{equation}
where $V_S$ is the pairing constant.
Within mean field theory, Eq.~\eqref{eq:BCS} can be expressed in the helical basis,
\begin{equation}
H_S=\frac12\sum_{\bf k,q,\lambda}t_{\lambda}({\bf k})\Delta({\bf q)}c^\dagger_{{\bf k}\lambda} c^\dagger_{{\bf -k+q}\lambda}+{\rm h.c.},
\end{equation}
where $t_{\lambda}({\bf k})=i\lambda  e^{-i\varphi_{\bf k}}$, while the order parameter solves the self-consistency equation
\begin{equation}
\label{eq:gap}
\Delta({\bf q)}=-\frac 1 2 V_S \sum_{{\bf k}\lambda}t_{\lambda}({\bf k})^*\langle c_{{\bf -k+q}\lambda} c_{{\bf k}\lambda}\rangle.
\end{equation}
Finally, potential scattering by impurities is accounted for by the term
\begin{equation}
H_{\rm imp}=\sum_{\bf k,k'}a_{\bf k}^\dagger U_0({\bf k\!-\!k'})a_{\bf k'}
=\sum_{\bf k,k'}c_{\bf k}^\dagger U_0({\bf k\!-\!k'})W_{\bf k,k'}c_{\bf k'},
\end{equation}
where $W_{\bf k,k'}=U^\dagger_{\bf k}U_{\bf k'}$ and the disorder potential satisfies the average
$\langle U_0({\bf q})U_0(-{\bf q'})\rangle_{\rm dis.}=U_0^2\delta_{{\bf q},{\bf q'}}$.

Introducing annihilation operators ${\cal C}_{{\bf k}\lambda}=(c_{{\bf k}\lambda},t_\lambda({\bf k})c^\dagger_{-{\bf k}\lambda})^T$ in particle-hole space, the total Hamiltonian takes the form $H=\frac12\sum_{{\bf k},{\bf k'}}{\cal C}_{\bf k}^\dagger{\cal H}_{\bf k,k'}{\cal C}_{\bf k'}$ with ${\cal C}_{\bf k}=({\cal C}_{{\bf k}+}^T,{\cal C}_{{\bf k}-}^T)^T$ and the Bogoliubov-de~Gennes Hamiltonian
\begin{eqnarray}
{\cal H}_{\bf k,k'}&=&
\left[\left(\xi_{\bf k}+\alpha k\sigma_z\right)\tau_z
+{\bf h}_{\bf k}\cdot{\bm \sigma}
\right]\delta_{\bf k,k'}
\\
&&
+U_0({\bf k-k'})W_{\bf k,k'}\tau_z
+\check\Delta({\bf k-k'}).\nonumber
\end{eqnarray}
Here, $\check\Delta({\bf k-k'})=\Delta({\bf k-k'})\tau_++\Delta^*({\bf k-k'})\tau_-$ and $\tau_i$ are Pauli matrices acting in particle-hole space with $\tau_\pm=(\tau_x\pm i\tau_y)/2$.

We now define the Matsubara Green's function ${\cal G}$ which solves the equation $(i\omega-{\cal H}){\cal G}=1$, where $\omega=(2n+1)\pi T$ ($n$ integer) is a Matsubara frequency at temperature $T$. Treating the disorder within the self-consistent Born approximation allows one to relate the impurity-induced self-energy to the disorder-averaged Green function $G=\langle {\cal G}\rangle_{\rm dis.}$. In a mixed momentum/space representation, this relation reads 
\begin{eqnarray}
\label{eq:self}
\Sigma_{\bf k}^{{\rm imp}}({\bf r})&=&U_0^2\tau_z\sum_{\bf k'} W_{\bf k,k'} G_{\bf k'}({\bf r})W_{\bf k',k} \tau_z,
\end{eqnarray}
provided that the Fermi wavelength is much smaller than the scale for spatial variations of $G_{\bf k}({\bf r})$. The gap equation \eqref{eq:gap} yields
\begin{equation}
\label{eq:gap2}
\Delta({\bf r)}=-\frac12 V_ST \sum_{\omega,{\bf k}}{\rm Tr}\left[\tau_-G_{\bf k}({\bf r})\right].
\end{equation}

The Dyson equation for the Green function $G_{\bf k}({\bf r})$ can be simplified within a quasiclassical approximation that assumes a separation between the large energy scales characterizing the electron spectrum in the normal state and the small energy scales associated with the superconducting properties. We will first derive the quasiclassical equations for the full matrix Green function. Importantly this requires to neglect the difference between the densities of states in the two helical bands. Thus, these equations are valid only in the case of a small spin-orbit coupling, $\alpha\ll v$, but at arbitrary disorder. We will then discuss how to take into account the difference between the densities of states to describe the helical phase at strong spin-orbit coupling, $\alpha\lesssim v$,  and moderate disorder, $1/\tau\ll\Delta_{\rm so}$.

\section{Weak spin-orbit coupling}

In the regime of weak spin-orbit coupling, $\alpha \ll v$, the splitting of the Fermi surface between different helical bands is negligible. In that case, the derivation of the quasiclassical Eilenberger equations solved by the energy-averaged Green function,
\begin{equation}
g({\bf r},\hat{\bf k})=\frac i\pi \int d\xi_{\bf k}\;\tau_z G_{\bf k}({\bf r}),
\end{equation}
is standard. We obtain
\begin{eqnarray}
\label{eq:eilenberger}
-v\hat{\bf k}\cdot{\bm \nabla}g({\bf r},\hat{\bf k})
&=&
\left[
\left(\omega+i\check\Delta({\bf r})+i{\bf h}_{\bf k}\cdot{\bm \sigma}\right)\tau_z
+i\Delta_{\rm so}\sigma_z
\right.
\nonumber\\
&&\left.
+\sigma^{\rm imp}({\bf r},\hat{\bf k}),g({\bf r},\hat{\bf k})\right],
\end{eqnarray}
with the reduced self-energy 
\begin{equation}
\label{eq:self-quasi}
\sigma^{\rm imp}({\bf r},\hat{\bf k})=
\frac 1{2\tau}U^\dagger_{\bf k}\langle U_{\bf k'} g({\bf r},\hat{\bf k}')U^\dagger_{\bf k'}\rangle_{\bf \hat k'}U_{\bf k}
.
\end{equation}
Here $1/\tau=2\pi \nu U_0^2 $ and $\langle \dots \rangle_{\bf \hat k}$ denotes averaging over the Fermi surface.
Using the explicit form of the matrices $U_{\bf k}$, 
one may write Eq.~\eqref{eq:self-quasi} as
\begin{eqnarray}
\sigma^{\rm imp}({\bf r},\hat{\bf k})&=&
\frac 1{4\tau}\Big\{
\langle g({\bf r},\hat{\bf k'})\rangle_{\hat{\bf k'}}+\sigma_x\langle g({\bf r},\hat{\bf k'})\rangle_{\hat{\bf k'}}\sigma_x
\\
&&-\hat {\bf k}\cdot\left((\sigma_x+
i\hat{\bf z}\times)[\langle \hat{\bf k'} g({\bf r},\hat{\bf k'})\rangle_{ \hat{\bf k'}},\sigma_x]\right)
\Big\}.\nonumber
\end{eqnarray}
The self-consistency equation \eqref{eq:gap2} reads
\begin{equation}\label{eq:gap3}
\Delta({\bf r)}=\frac {i\pi V_S \nu} 2 T\sum_\omega \mathrm{Tr} \left[\tau_- \langle g({\bf r},\hat{\bf k})\rangle_{\hat{\bf k}}\right].
\end{equation}
Moreover, the normalization condition $g^2({\bf r},\hat{\bf k})=1$ must be satisfied.

At large disorder, $1/\tau\gg h,\Delta,T$, the quasiclassical Green function is essentially isotropic over the Fermi surface. That is, we may approximate $g({\bf r},\hat{\bf k})\approx \bar g({\bf r})+ \hat{\bf k}\cdot\bm{g}({\bf r})$, where the second term is a small anisotropic correction. The normalization condition imposes $\bar g^2=1$ and $\{\bar g,\bm{g}\}=0$. If, furthermore, we assume  $\Delta_{\rm so}\gg h,\Delta,T$, Eq.~\eqref{eq:eilenberger} averaged over the Fermi surface yields
\begin{equation}
\label{eq:quasi-av0}
\left[i\Delta_{\rm so}\sigma_z+\frac1{4\tau}(\bar g +\sigma_x \bar g \sigma_x),\bar g\right]=0,
\end{equation} 
 in  leading order in $1/\tau$ and $\Delta_{\rm so}$. Thus, it is satisfied if $\bar g({\bf r})= g_0({\bf r})+\delta g({\bf r})$, with $[\sigma_x,g_0({\bf r})]=[\sigma_z,g_0({\bf r})]=0$ and $\delta g({\bf r})\ll g_0({\bf r})$. 
 
We therefore look for a solution $g_0({\bf r})$ that is diagonal in helical space and obeys the normalization conditions $g_0^2=1$ and $\{g_0,\bm{g}\}=\{g_0,\delta g\}=0$. As a first step, we determine $\bm{g}$ by inserting the expansion for $g({\bf r},\hat{\bf k})$ into Eq.~\eqref{eq:eilenberger} and averaging over the Fermi surface after multiplication with $\hat {\bf k}$. This yields
\begin{eqnarray}
\label{eq:quasi-av1}
&&-v\bm{\nabla}g_0-[(\sigma_x
+i\hat{\bf z}\times){\bf h}_\parallel\sigma_z\tau_z,g_0]
\\
&=&[i\Delta_{\rm so}\sigma_z+\frac 1{2\tau}g_0,\bm{g}]
-\frac1{8\tau}
[(\sigma_x
+i\hat{\bf z}\times)[\bm{g},\sigma_x],g_0].
\nonumber
\end{eqnarray} 
Using $\{g,\bm{g}\}=0$, we find that Eq.~\eqref{eq:quasi-av1} is solved with 
\begin{eqnarray}
\label{eq:corr-g-anis}
\bm{g}&=&-v\tau g_0\bm{\nabla}g_0+2i\tau\left(1\!+\!\frac1{4\tau^2\Delta_{\rm so}^2}\right)
({\bf h}_\parallel\times\hat{\bf z})\sigma_zg_0[\tau_z,g_0]\nonumber\\
&& -i\frac{{\bf h}_\parallel}{\Delta_{\rm so}}\left(\sigma_x-\frac1{2\tau\Delta_{\rm so}}
\sigma_yg_0\right)[\tau_z,g_0].
\end{eqnarray}
As a second step, we find the isotropic correction $\delta g$ by considering the terms in the next-order correction to Eq.~\eqref{eq:quasi-av0} that are non-diagonal in helical space. 
We obtain
\begin{eqnarray}
\label{eq:corr-g-iso}
\delta g&=&-i\frac{h_z}{2\Delta_{\rm so}}\left(\sigma_y+\frac1{2\tau\Delta_{\rm so}}
\sigma_xg_0\right)[\tau_z,g_0].
\end{eqnarray}
Finally, taking  into account Eqs.~\eqref{eq:corr-g-anis} and \eqref{eq:corr-g-iso} to obtain the  diagonal terms,
we find that $g_0$ obeys a standard Usadel equation,
\begin{eqnarray}
\label{eq:usadel}
D\bm{\nabla}\cdot (g_0\bm{\nabla} g_0)=\left[(\omega+i\check\Delta)\tau_z+\frac\Gamma2\tau_zg_0\tau_z,g_0\right],
\end{eqnarray} 
where $D=\mu\tau/m$ is the diffusion constant and 
\begin{equation}
\Gamma=2\tau h_\parallel^2\left(1+\frac1{2\tau^2\Delta^2_{\rm so}}\right)+\frac{h_z^2}{2\tau\Delta_{\rm so}^2}\label{eq-gamma}
\end{equation}
is an effective pair-breaking parameter. Note that the orbital effect of the magnetic field may be readily accounted for in Eq.~\eqref{eq:usadel} with the substitution $\bm{\nabla}\to\bm{\nabla}-ie{\bf A}[\tau_z,.]$, where $\bf A$ is the vector potential.

Properties of superconductors described by Eq.~\eqref{eq:usadel} are well established~\cite{ag}. In the absence of a magnetic field $\Gamma=0$, and therefore the superconducting properties are insensitive to the spin-orbit coupling. A finite magnetic field suppresses superconductivity. At $T=0$, a second-order transition to the normal state occurs at the upper critical field determined by 
the equation $\Gamma+\Gamma_{\rm orb}=\Delta_0/2$, where $\Gamma_{\rm orb}=eDB_z$ and $\Delta_0$ is the superconducting gap at zero temperature in the absence of the magnetic field. 
Thus, the in-plane critical field is given as
\begin{eqnarray}
h_{\parallel c}&=&\frac {\Delta_{\rm so}}2\sqrt{\frac{2\Delta_0\tau}{1+2\tau^2\Delta_{\rm so}^2}}
\approx
\begin{cases}
\frac12\sqrt{\frac{\Delta_0}\tau},&\frac1\tau\ll\Delta_{\rm so},\\
\Delta_{\rm so}\sqrt{\frac{\Delta_0\tau}2},&\frac1\tau\gg\Delta_{\rm so}.
\end{cases}
\nonumber\\
\end{eqnarray}
In the out-of-plane configuration, the paramagnetic mechanism yields a critical field
 $h^{\rm p}_{\perp c}=\Delta_{\rm so}\sqrt{\Delta_0\tau}$, which is much larger than the in-plane critical field $h_{\parallel c}$ for $1/\tau\ll\Delta_{\rm so}$. However, typically the out-of-plane critical field is determined by the orbital mechanism yielding $h^{\rm orb}_{\perp c}=\frac12 g\mu_B\phi_0/(2\pi\xi^2)$, where $\phi_0$ is the flux quantum and $\xi=\sqrt{D/\Delta_0}$ is the superconducting coherence length. 

Using these results, we can now check our assumptions used in the derivation of the Usadel equation. We find that the approximations ${\bm g},\delta g\ll g_0$ are well obeyed 
provided that $\Delta_0\ll 1/\tau\ll\Delta_{\rm so}^2/\Delta_0$. To describe the system at larger disorder, $1/ \tau\gg\Delta_{\rm so}^2/\Delta$, the spin basis~\cite{Bergeret2013}  is better suited than the helical basis. In the regime $\Delta_{\rm so}\ll 1/\tau\ll\Delta_{\rm so}^2/\Delta_0$, both approaches are valid and yield the same results. 

Further examining Eq.~\eqref{eq:corr-g-anis}, we note that, at $1/ \tau\ll\Delta_{\rm so}$, the anisotropic correction to $g$ is essentially diagonal in helical space.
This suggests that, in this regime, we may project the Green function onto the helical bands{, thereby neglecting the small inter-band superconducting correlations}. In doing so, we then are able to account for
{ both the large intra-band superconducting correlations and} the difference in the densities of states between the helical bands, and thus access the helical phase at strong spin-orbit coupling, $\alpha\lesssim v$.

\section{Strong spin-orbit coupling}

As shown above, at ``moderate'' disorder, $1/ \tau\ll\Delta_{\rm so}$, superconductivity takes place separately in each of the helical bands. In that case, we may treat the spin-orbit splitting non-perturbatively by deriving the quasiclassical equations for the Green functions projected onto the helical bands,
\begin{equation}
\label{eq:g-helical} 
g({\bf r},\hat{\bf k})\approx \left(\begin{array}{cc}
g_+({\bf r},\hat{\bf k}) & 0\\
0&g_-({\bf r},\hat{\bf k}) 
\end{array}
\right),
\end{equation}
where $g_\lambda({\bf r},\hat{\bf k})$ are matrices in particle-hole space only. Inserting the ansatz \eqref{eq:g-helical} into Eq.~\eqref{eq:self} and performing the quasiclassical approximation, we find
\begin{equation}
\label{eq:eilenberger-helical}
-v\hat{\bf k}\cdot{\bm \partial}_\lambda g_\lambda({\bf r},\hat{\bf k})
\!=\!
\left[
\left(\omega+i\check\Delta({\bf r})\right)\tau_z
+\sigma^{{\rm imp}}_\lambda({\bf r},\hat{\bf k}),g_\lambda({\bf r},\hat{\bf k})\right],
\end{equation}
where ${\bm\partial}_\lambda={\bm \nabla}-i(\lambda/v) ({\bm h}_\parallel\times\hat {\bf z})[\tau_z,.]$ and 
\begin{eqnarray}
\label{eq:self-quasi-helical}
\sigma^{{\rm imp}}_\lambda({\bf r},\hat{\bf k})&=&\sum_{\lambda'}
\frac 1{4\tau_{\lambda'}}\Big(\langle g_{\lambda'}({\bf r},\hat{\bf k'})\rangle_{\hat{\bf k'}}\\
&&
\qquad\qquad+\lambda\lambda'\hat {\bf k}\cdot \langle \hat{\bf k'} g_{\lambda'}({\bf r},\hat{\bf k'})\rangle_{\hat{\bf k'}}
\Big).\nonumber
\end{eqnarray}
Here, the different densities of states in the helical bands result in different scattering rates $1/\tau_\lambda=2\pi \nu_\lambda U_0^2$.
The different densities of states also appear in the self-consistency equation \eqref{eq:gap2}  that now reads
\begin{equation}
\label{eq:gap-helical}
\Delta({\bf r)}=\frac {i\pi V_S } 2 T\sum_{\omega,\lambda} \nu_\lambda\, \mathrm{tr} \left[\tau_- \langle g_\lambda({\bf r},\hat{\bf k})\rangle_{\hat{\bf k}}\right].
\end{equation}
As usual, the normalization condition $g_\lambda^2({\bf r},\hat{\bf k})=1$ must be satisfied.

In the presence of disorder, we proceed as in the case of weak spin-orbit coupling. Namely, we approximate $g_\lambda({\bf r},\hat{\bf k})\approx \bar g_\lambda({\bf r})+\hat{\bf k}\cdot {\bm g}_\lambda({\bf r})$ and $\bar g_\lambda({\bf r})\approx g_0({\bf r})+\lambda \delta g({\bf r})$, with $g_0^2=1$ and $\{g_0, \delta g\}=\{g_0, {\bm g}_\lambda\}=0$. Inserting this expansion into Eq.~\eqref{eq:eilenberger-helical} and performing the averages over $\hat{\bf k}$ as discussed above, we find
\begin{eqnarray}
{\bm g}_\pm&=&-v\tau^2g_0 \left(\frac1{\tau_\pm}{\bm \partial}_\pm g_0 
\pm i \frac1{v\tau_\mp}({\bm h}_\parallel\times \hat{\bf z})[ \tau_z,g_0]
\right)\!,\quad
\\
\delta  g&=&-\frac{v^2\tau}{4\alpha}g_0\left({\bm \partial}_+\cdot {\bm g}_+-{\bm \partial}_-\cdot {\bm g}_-\right),
\end{eqnarray}
where $1/\tau=\sum_{\lambda=\pm}1/(2\tau_\lambda)$. Finally, we obtain the Usadel equation
\begin{equation}
\label{eq:usadel2}
\tilde D\tilde{\bm{\partial}}\cdot (g_0\tilde{\bm{\partial}} g_0)=\left[(\omega+i\check\Delta)\tau_z+\frac{\tilde \Gamma} 2\tau_z g_0
\tau_z,g_0\right],
\end{equation}
where $\tilde D=\tau(v^2+\alpha^2)/2$, $\tilde{\bm \partial}={\bm \nabla}+2i\alpha/(\alpha^2+v^2)({\bm h}_\parallel\times \hat{\bf z})[\tau_z,.]$, and $\tilde \Gamma=2 \tau h_\parallel^2(v^2-\alpha^2)/(v^2+\alpha^2)$.
Equation \eqref{eq:usadel2} which describes Rashba superconductors in the regime $\Delta\ll 1/\tau\ll\Delta_{\rm so}$ is the main result of our work.

In addition to a pair-breaking parameter, $\tilde\Gamma$, which reduces to Eq.~\eqref{eq-gamma} in the limit $\alpha\ll v$ and $1/\tau\ll\Delta_{\rm so}$, the Usadel equation \eqref{eq:usadel2} contains an effective vector potential.
This vector potential may be eliminated by the unitary transformation $\tilde g_0({\bf r})=V({\bf r})g_0({\bf r})V^\dagger({\bf r})$, where $V({\bf r})=\exp[-(i/2){\bf q\cdot r}\,\tau_z]$ and
\begin{equation}
{\bf q}=-\frac{4\alpha}{\alpha^2+v^2}({\bm h}_\parallel\times \hat{\bf z}).
\end{equation}
Under this transformation, the order parameter transforms as $\Delta({\bf r})\to \Delta({\bf r}) \exp[-i{\bf q\cdot r}]=const$, which corresponds to a spatial modulation. This is the so-called helical phase. The modulation is necessary to realize the ground state with zero current~\cite{Dimitrova2003}. Furthermore, it leads to an increase in the upper critical field
\begin{equation}
h_{\parallel c}^{({\bf q})}=\frac12\sqrt{\frac{\Delta_0}\tau\frac{v^2+\alpha^2}{v^2-\alpha^2}}.
\end{equation}
Note that the out-of plane component of the magnetic field does not contribute after projecting the Green functions onto the helical bands~\cite{footnote2} .

\section{Ginzburg-Landau functional}

\begin{figure}
\includegraphics[width=0.9\columnwidth]{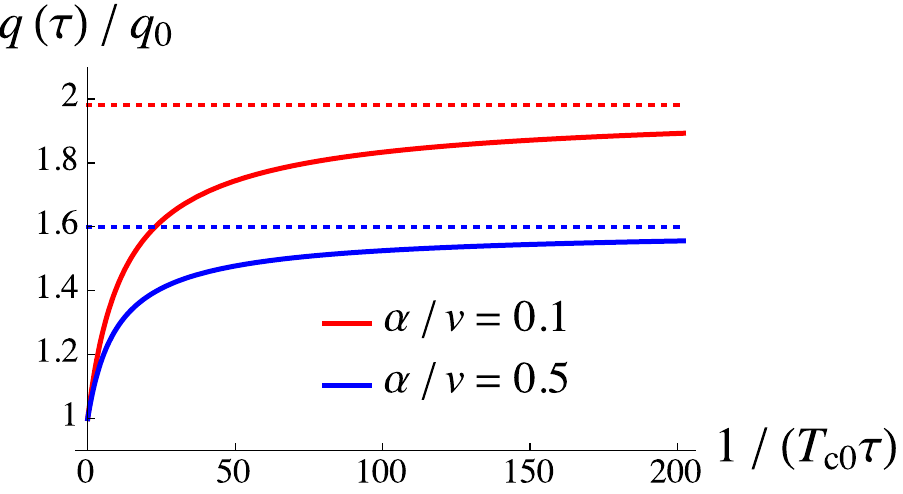}
\caption{\label{fig}  Helical modulation wavevector [in units of $q_0=2\alpha h_\parallel/v^2$] as a function of the disorder strength, for various amplitudes of the spin-orbit coupling, near the superconducting critical temperature.} 
\end{figure}

Near the second-order transition to the normal state, the order parameter vanishes. In that regime, it is sufficient to solve the Eilenberger equation \eqref{eq:eilenberger-helical} perturbatively in $\Delta$ around the normal state solution, $g^{(0)}_\lambda({\bf r},\hat{\bf k})=\tau_z\,{\rm sign}(\omega)$. Inserting the perturbative solution up to the third order into the gap equation \eqref{eq:gap-helical}, we obtain as the  result of a long, but straightforward calculation
\begin{widetext}
\begin{eqnarray}
\frac 1 {\nu V_S}\Delta({\bf q})=2\pi T\sum_{\omega>0}
\Bigg\{\left[
\frac1\omega-\frac{h_{\parallel}^2}{\omega^2(\omega+\Omega)}
-\frac{v^2q^2}{8\omega^2\Omega}\left(1+\frac{\alpha^2}{v^2}\frac1{2\tau(\omega+\Omega)}\right)-\frac\alpha{\omega^2(\omega+\Omega)}{\bf q}\cdot({\bf h}_\parallel\times \hat{\bf z})
\right]
\Delta({\bf q})
\nonumber\\
-
\frac1{2\omega^3}\sum_{{\bf q}_1+{\bf q}_2-{\bf q}_3={\bf q}}\Delta({\bf q}_1)\Delta({\bf q}_2)\Delta^*({\bf q}_3)
\Bigg\},
\end{eqnarray}
assuming $h,vq\ll1/\tau,T\ll\Delta_{\rm so}$. Here we used notation $\Omega=\omega+1/(2\tau)$.
For temperatures close to the critical temperature, this equation minimizes the free energy functional
\begin{eqnarray}
{\cal F}=\nu\int (d^2r)
\left\{\left(
\frac{T-T_{c0}}{T_{c0}}+\gamma h_\parallel^2\right)|\Delta|^2
+\frac{7\zeta(3)}{16\pi^2T_{c0}^2}|\Delta|^4
+\beta|{\bm \nabla}\Delta|^2
-\frac i 2\kappa\left({\bf h}_\parallel\times \hat{\bf z}\right)\left(\Delta^*{\bm \nabla}\Delta-\Delta{\bm \nabla}\Delta^*\right)
\right\},
\end{eqnarray}
\end{widetext}
where $T_{c0}$ is the critical temperature at zero magnetic field, $\beta=(\pi T v^2/4)\sum_{\omega>0}[1+\alpha^2/(v^2(1+4\omega\tau))]/(\omega^2\Omega)$, and $\kappa=\alpha\gamma=2\alpha\pi T\sum_{\omega>0}1/[\omega^2\left(\omega+\Omega\right)]$. The modulation of the helical phase is easily obtained as
${\bf q}=-\kappa/(2\beta){\bf h}_\parallel \times \hat{\bf z}${; its dependence on the disorder strength is shown in Fig.~\ref{fig}}. In particular, this result confirms that, in the lowest order in $\alpha/v$,
the helical modulation wavevector only varies by a factor 2 between the clean~\cite{Edelstein1996,Barzykin2002} and dirty limit~\cite{Dimitrova2003}. Note that the critical temperature is shifted downward, $T_c(h_\parallel)=T_{c0}(1-\gamma h_\parallel^2+\beta q^2)$. At any disorder, the shift vanishes when $\alpha\to v$ (a result that was found previously in the clean case~\cite{MineevSigrist}){, while the helical modulation wavevector remains constant}.

\section{Conclusions}

In this work, we showed that the helical phase that arises in two-dimensional Rashba superconductors can be described with quasiclassical equations, both in the clean and { moderately} disordered case. Importantly, the difference between the densities of states in the helical bands can be accounted for within quasiclassics because, even though { impurities}
lead to inter-band scattering, superconductivity remains essentially intra-band
{ when the disorder-induced broadening of the helical bands does not exceed their energy splitting}. 

Beyond the bulk properties discussed here, a natural application of this formalism will be to investigate the proximity effect between a conventional superconductor and two-dimensional semiconductors with a large spin-orbit coupling such as, for example, InAs. Furthermore, it will be of interest to generalize the formalism to different types of spin-orbit coupling as well as to the three-dimensional case.

\acknowledgements
We thank Vladimir Mineev for valuable discussions. Furthermore, we acknowledge support by ANR through
grants ANR-11-JS04-003-01 and ANR-12-BS04-0016-03,
and an EU-FP7 Marie Curie IRG.

\end{document}